\documentclass[12pt,a4paper]{article}
\usepackage{amsmath}
\usepackage{amsfonts}
\usepackage{amssymb}
\usepackage{natbib}
\usepackage{algorithm}
\usepackage{algpseudocode}
\usepackage{amsthm}
\usepackage{graphicx}
\usepackage{mathtools}
\usepackage[utf8]{inputenc}
\DeclarePairedDelimiter{\norm}{\lVert}{\rVert}

\numberwithin{equation}{section}

\begin{document}

\begin{titlepage}
   \vspace*{\stretch{1.0}}
   \begin{center}
      \Large\textbf{Sparse Non-Convex Optimization For Higher Moment Portfolio Management}\\
	  \large\textit{Farshad Noravesh}\footnote{Email: noraveshfarshad@gmail.com}

   \end{center}
   \vspace*{\stretch{2.0}}
\begin{abstract}
One of the reasons that higher order moment portfolio optimization methods are not fully used by practitioners in investment decisions is the complexity that these higher moments create by making the optimization problem nonconvex. Many few methods and theoretical results exists in the literature, but the present paper uses the method of successive convex approximation for the mean-variance-skewness problem.\\
\textbf{Keywords}: nonconvex, $\ell_{1}$ regularization, higher moment, portfolio optimization, portfolio selection, skewness
\end{abstract}

\end{titlepage}

\section{Introduction}
Although Modern portfolio theory started by Harry Markowitz by the classical mean-variance optimization, there are many shortcoming on this theory from different perspectives. The first shortcoming is that it does not care about higher moments which becomes a critical issure when assets are highly skewed and having big kurtosis which is far from normal distribution. One of the obstacles for development of higher order moment portfolio optimization methods is the nonconvexity that is generated by considering skewness and kurtosis. There are four major approaches to handles nonconvex optimization problems. The first groups of methods is based on stochastic gradient descent and its convergence rates are far from other methods as is described in  \citep{lei2020}. The second approach is to see noncovex functions as a difference of some convex functions but this framework does not work for any class of nonconvex problems. The third approach is to use the idea of majorization minimization (MM) by iteratively minimize a surrogate function that upper-bounds the objective function but has the same derivative with it at the current iteration which is explained in \citep{Sun2017}. The forth approach is called successive convex approximation and is used in the present paper to show practitioners how to use it for mean-variance-skewness problem. This method is very similar to MM method but requires the upper bound function to be a convex function.

The second shortcoming is the uncertainty around the parameters inside mean return, covariance and coskewness which also makes the portfolio solutions very unstable. Also if incorrect parameters are used the solution is only suboptimal. There are three major approaches to tackle this problem. The first approach comes from frequentists that try to estimate the covariance matrices and others using maximum likelihood or any other frequentist approaches. One may argue that simple averaging could help, but since historical data are being used, only recent data can be relevant which reduces the overall data that are taken into considerations and creates big volatilities. The second approach is the bayesian approach which utilizes some intrinsic structure such as Fama-French three-factor model as a prior to shrink the mean and covariance toward the structure. The third approach which is of interest in the present paper is penalizing the objective function such as \citep{Ho2015} which uses the weighted elastic net for penalizing the mean-variance portfolio. There are huge literature on the fact that why these  $ \ell_{1}$-norm or  $ \ell_{2}$-norm or any combination of them is helpful. For example \citep{Brodie2009} mentioned four reasons on the importance of $ \ell_{1}$-norm. Apart from the fact that sparsity limits the number of positions that must be monitored and liquidated, they noticed that it reduce the sensitivity of the optimization to the possible collinearities between the assets and therefore has good stability and this is possible with using only limited amount of training data.\citep{Fan2012} showed the following important result which bounds the risk minimization error by $ \ell_{1}$-norm.
\begin{equation}
|R(w,\hat{\Sigma})-R(w,\Sigma)|\leq\norm{\hat{\Sigma}-\Sigma}_{\infty}\norm{w}_{1}^{2}
\end{equation}
where risk is defined by $\Sigma=w^{T}\Sigma w$

\section{MVS portfolio}
Mean-Variance-Skewness portfolio selection and optimization can be defined as solution of \\ 
\begin{equation} \label{eq-sparseMVS}
\underset{w}{minimize} \ f(w)+g(w)=-\lambda_{1}\phi_{1}(w)+\lambda_{2}\phi_{2}(w)-\lambda_{3}\phi_{3}(w)+\lambda_{4}\norm{w}_{1} 
\end{equation}
where $\phi_{1}$,$\phi_{2}$,$\phi_{3}$ are mean,variance and skewness respectively and are defined as follows:
\begin{equation} \label{eq-MVS}
\begin{split}
\phi_{1}(w)&=w^{T}\mu  \\
\phi_{2}(w)&=w^{T}\Sigma w  \\
\phi_{3}(w)&=w^{T}\Phi(w\otimes w) 
\end{split}
\end{equation}
where $\Phi$ is the co-skewness matrix. As is shown in \citep{Zhou2021}, the gradient and hessian of skewness can be written as:
\begin{equation} \label{eq-gradientAndHessian}
\begin{split}
\nabla\phi_{3}&=3\Phi(w\otimes w)  \\
\nabla^{2}\phi_{3}&=6\Phi(I\otimes w)
\end{split}
\end{equation}
\citep{Byrd2013} could be used to solve the optimization problem in \eqref{eq-sparseMVS} but that algorithm has big computational complexity since it uses newton method and in general second order methods are harder than first order methods although they could be more accurate in some algorithms. Therefore an algorithm is developed in the next section that is computationally easier and also much simpler to implement. 
\section{Problem Formulation}
If $\lambda_{3}$ in \eqref{eq-sparseMVS} is zero, then \citep{Kremer2020} has used the  $ \ell_{1}$-norm regularization to do sparse portfolio selection but skewness and other higher moments are not handled in their article. If $\lambda_{4}$ in \eqref{eq-sparseMVS} is zero, then successive convex approximation method can be used as is described in \citep{Zhou2021}, otherwise the optimization functions can be rearranged like \citep{Yang2017} to handle nondifferentiability of the $ \ell_{1}$ penalizing term.
\eqref{eq-sparseMVS} can be written as 
\begin{equation} \label{eq-yang}
\begin{split}
\underset{w,y}{minimize} \ & f(w)+y \\
subject \ to \ & g(w)\leq y
\end{split}
\end{equation}
where $g=\lambda_{4}\norm{w}_{1}$. Since f is a nonconvex function, successfive psudoconvex approximation can be used to sequentially solve minimization of a set of surogate approximation functions. Psuedoconvex surogate functions can be used as explained in \citep{Yang2017} to broaden the type of approximate functions that can be chosen. Let $\tilde{f}(w;w^t)$ be the approximation of f at iteration t. Thus, the solution to approximate function of problem in \eqref{eq-yang} around $(w^{t},y^{t})$ is
\begin{equation}\label{eq-approximateProblem}
(Bw^{t},y^{\star}(w^t))=\underset{(w,y):w\in W,g(w)\leq y}{arg min} \  \tilde{f}(w;w^{t})+y  
\end{equation}
The step size $\gamma^{t}$ can be easily achieved by the following exact line search algorithm.
\begin{equation}
\gamma^{t}\in \underset{0\leq\gamma\leq1}{argmin}\{f(w^{t}+\gamma(Bw^{t}-w^{t})+y^{t}+\gamma(y^{\star}(w^{t})-y^{t}) )   \}
\end{equation}
where $y^{t}\geq g(x^{t})$. Now the update can be calculated as follows
\begin{equation} \label{eq-updateYang}
\begin{split}
w^{t+1}&=w^{t}+\gamma^{t}(Bw^{t}-w^{t})  \\
y^{t+1}&=y^{t}+\gamma^{t}(y^{\star}(w^{t})-y^{t})
\end{split}
\end{equation}
By substituting \eqref{eq-MVS} in \eqref{eq-sparseMVS} and picking the differentiable part, the following nonconvex function results
\begin{equation} \label{eq-originalDifferentiable}
f(w)=\underbrace{-\lambda_{1}w^{T}\mu+\lambda_{2}\omega^{T}\Sigma\omega}_{\text{convex}}-\underbrace{\lambda_{3}w^{T}\Phi(w\otimes w)}_{\text{non-convex}}
\end{equation}
The first two terms in \eqref{eq-originalDifferentiable} are convex functions and therefore there is no need to approximate them by convex functions, the third term which is the portfolio skewness should be approximated by a second order function as 
\begin{equation}\label{eq-parametricApprox}
\begin{split}
\tilde{f}_{ncvx}(w,w^{t})=f_{ncvx}(w^{t})+\nabla f_{ncvx}(w^{t})^{T}(w-w^{t}) & \\
+ \frac{1}{2}(w-w^{t})^{T}\nabla^{2}f_{ncvx}^{t}(w^{t})(w-w^{t})
\end{split}
\end{equation}
Approximating the nonconvex part of \eqref{eq-originalDifferentiable} by \eqref{eq-parametricApprox} yields
\begin{equation}
\begin{split}
\tilde{f}_{ncvx}=-\lambda_{3}(w^{t})^{T}\Phi(w^{t}\otimes w^{t})+3\Phi(w^{t}\otimes w^{t})^{T}(w-w^{t}) \\
+\frac{1}{2}(w-w^{t})^{T}(-\lambda_{3}(6\Phi(I\otimes w^{t})))(w-w^{t})
\end{split}
\end{equation}
Thus, the solution to the following constraint convex optimization problem is used to update \eqref{eq-updateYang} for the next update which generates a point for another successive approximation. 
\begin{equation} \label{eq-practical}
\begin{split}
\underset{w}{minimize} & -\lambda_{1}w^{T}\mu+\lambda_{2}w^{T}\Sigma w+\tilde{f}_{ncvx}  \\
subject \ to \ &  \norm{w}_{1}\leq y
\end{split}
\end{equation}
\eqref{eq-practical} can be solved by convex optimization methods such as \citep{Solntsev2015} which uses a proximal gradient method since proximal operator for  $ \ell_{1}$-norm is easy to evaluate.
The full algorithm in the present paper is summarized in Algorithm~\ref{alg:nonconvexSparse}
\begin{algorithm}[H]
\caption{sparse nonconvex optimization for mean-variance-skewness}
\label{alg:nonconvexSparse}
Input: given returns of a portfolio having N assets \\
1: calculate mean,variance and coskewness matrices  \\
2: initialize the weight vector randomly \\
Loop: until the local minimum of approximate function converges to local minimum of original function. \\
3: substitute weights and calculate variables using \eqref{eq-gradientAndHessian} \\
4: solve \eqref{eq-practical} to find $Bw^t$ \\
5: update the weights and y from \eqref{eq-updateYang}
\end{algorithm}

\begin{figure}[H]
  \includegraphics[width=\linewidth]{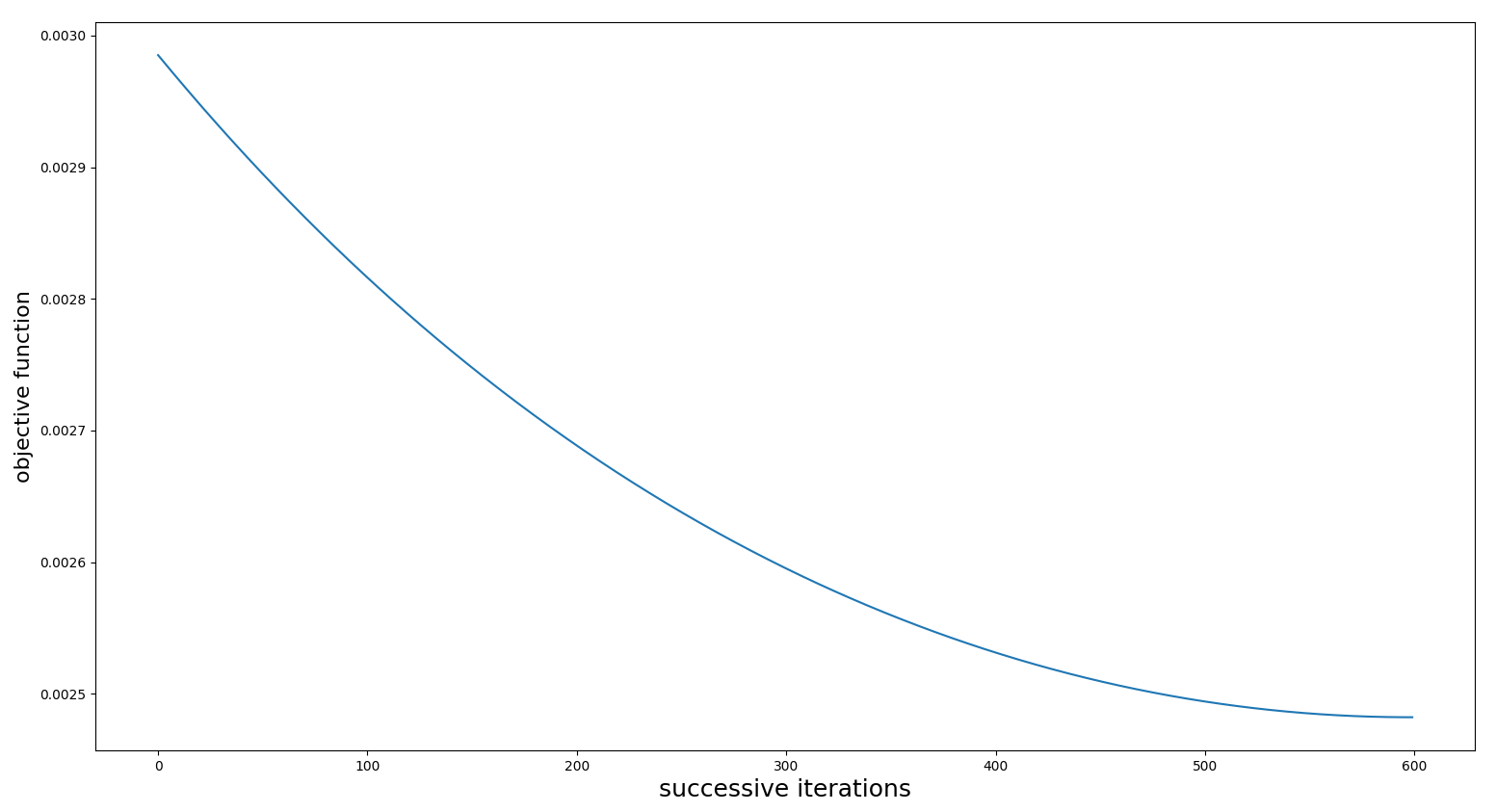}
  \caption{objective value in terms of successive iterations}
  \label{fig:function}
\end{figure}
\begin{figure}[H]
  \includegraphics[width=\linewidth]{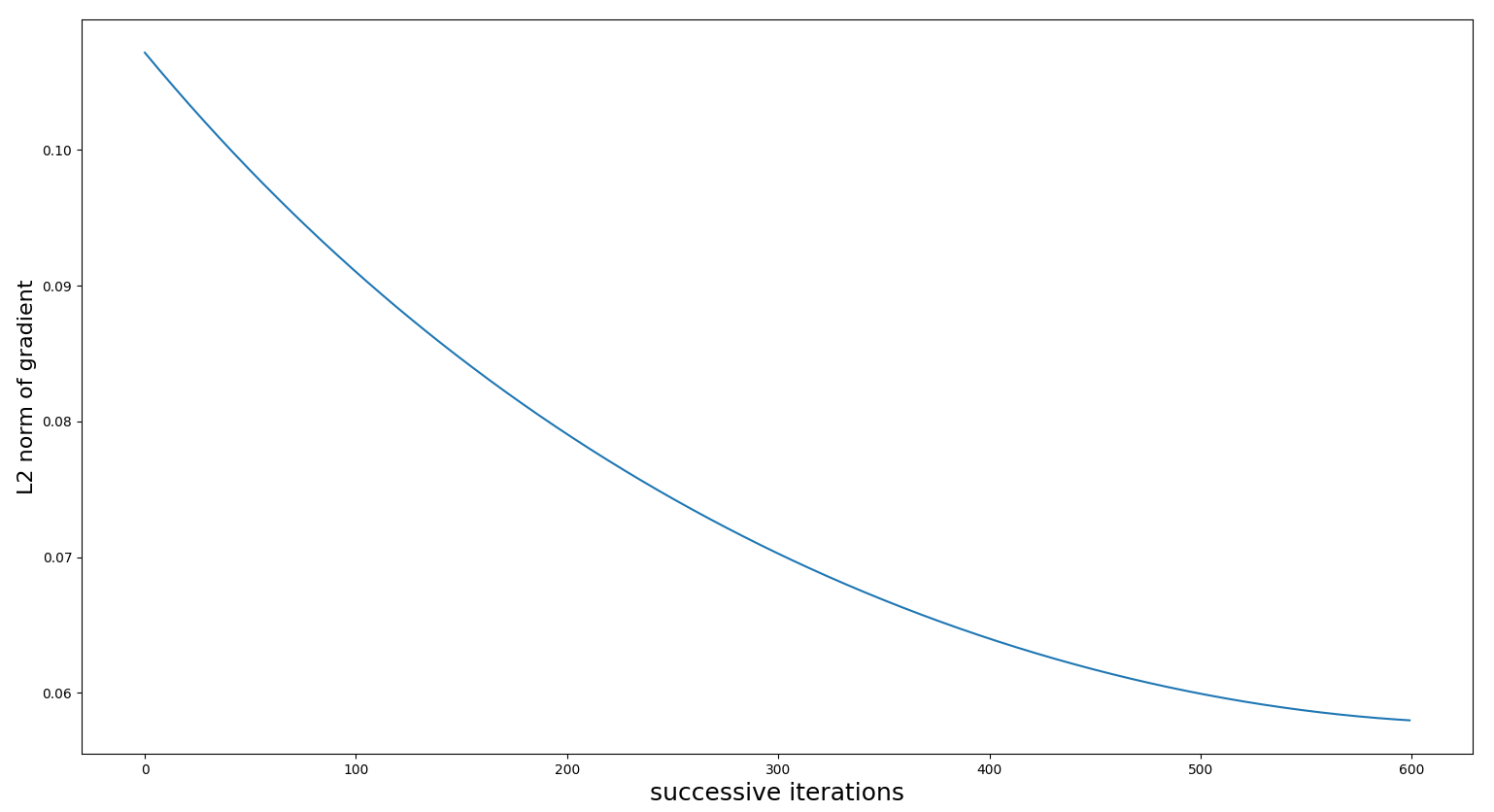}
  \caption{L2 norm of gradient of objective function in terms of successive iterations}
  \label{fig:gradientNorm}
\end{figure}

 
Figure~\ref{fig:function} shows that the value of function in \eqref{eq-sparseMVS} evaluated on weights at each iteration $t$  is reduced. The $ \ell_{2}$-norm of gradient is shown in Figure~\ref{fig:gradientNorm} and demonstrates how it decays as successive iterations increases.  

\bibliographystyle{agsm}
\bibliography{successiveConvex}
\end{document}